\documentclass[twocolumn,prl]{revtex4}

\newcommand{\BE}{\begin{equation}}
\newcommand{\EE}{\end{equation}}
\newcommand{\BA}{\begin{eqnarray}}
\newcommand{\EA}{\end{eqnarray}}
\usepackage{graphicx}
\usepackage{dcolumn}
\usepackage{bm}
\input epsf.tex

\begin{document}

\begin{flushleft}
{\bf Comment on ``Total Negative Refraction in Real Crystals for
Ballistic Electrons and Light"} (Phys. Rev. Lett. {\bf 91}, 157404
(2003))
\end{flushleft}

Recently, Zhang {\it et al.} \cite{Zhang} have demonstrated that
an amphoteric refraction, i.~e. both positive and negative
refraction, may prevail at the interface of two uniaxial
anisotropic crystals when their optical axes are in different
directions. The authors subsequently made a correspondence between
such a refraction with the negative refraction expected for Left
Handed Materials (LHMs). Here we comment that the amphoteric
refraction can be observed even with one uniaxial crystal, and the
refraction is {\it not} related to the negative refraction
expected for the much debated LHM. Rather, the phenomenon is a
natural result of anisotropic media.

Our experiments have revealed the amphoteric refraction with
45$^o$-cut calcite crystals in air\cite{You}. Fig.~1 is a
photograph of the experimental results showing both positive and
negative refractions. The amphoteric refraction is clearly
demonstrated. For the positive refraction shown in Fig.~1(a), the
incident and refraction angles are 25.9 and 10.0 degrees
respectively. In the case of the negative refraction (Fig.~1(b)),
it is seen that the refracted ray bends backward to the same side
of the incident light, and the incident angle and the refraction
angle are found to be 3.9$^o$ and -4.0$^o$, respectively.

The amphoteric refraction observed is purely due to the anisotropy
of refractive media, and can be well elucidated by considering the
normal surface of the wave vectors\cite{Yeh}. The essence is that
(1) the energy flow direction is characterized by the group
velocity given as $\vec{v}_g =
\nabla_{\vec{k}}\omega(\vec{k})$\cite{Yeh}; therefore the
direction of the energy flow, i.~e. the Poynting vector, will be
the normal to the wavefront surface; (2) the tangential components
of the wave vectors must be continuous across the interface
between two media. Based upon these, we have derived an equation
relating the incident angle in air and the refractive angle of the
energy ray as $ \sin\theta_i =
\frac{n_e^2\tan(\theta_s-\theta_c)\cos\theta_c +
n_o^2\sin\theta_c}{\sqrt{n_o^2 +
n_e^2\tan^2(\theta_s-\theta_c)}},$ where $n_o$ and $n_e$ are the
ordinary and extraordinary indices of the crystal, and $\theta_i$
and $\theta_s$ are the incident and refraction angles
respectively, while $\theta_c$ is the angle between the principal
optical axis and the interface normal. The analysis is illustrated
by Fig.~2. The experimental results in Fig.~1 fully agree with the
theory.

Although the refraction behavior of anisotropic crystals resembles
that of LHMs in some way, there are critical differences. In the
anisotropic crystal optics, the negative refraction occurs only in
a narrow range of incident angles. Accordingly, a slab of
anisotropic crystals will not exhibit the superlensing effect, a
unique feature expected for LHMs\cite{Pendry}.

\begin{flushleft}
Hon-Fai Yau$^{1*}$, Jung-Ping Liu$^1$, Bowen Ke$^1$, Chao-Hsien
Kuo$^2$, and Zhen Ye$^2$\\
{\small \it $^1$Institute of Optical Science, and $^2$Department
of Physics, National Central University, Chungli, Taiwan, ROC.}
\\ $^*$Email: yauhf@ios.ncu.edu.tw

\end{flushleft}

\section*{Figure Captions}

\begin{description}

\item[Figure 1] Amphoteric refraction of light beam observed for
calcite crystals in air (not to scale). The light source is the
He-Cd laser with wavelength 441.6 nm. The refraction indices along
the two principal axes of the crystals at this wavelength are
1.6741 and 1.4937 respectively (www.casix.com.)

\item[Figure 2]

\label{fig2}\small The mechanism for amphoteric refraction by
anisotropic media. The Poynting and wave vectors are denoted by
$\hat{S}$ and $\vec{k}$; the $k_y$ components should be continuous
across the interface. When the incidence is towards right, there
is a minimum angle $\phi_{min}$. When the incident angle is
greater than $\phi_{min}$, the refraction will be positive
(indicated by $\hat{S}_1'$, relative to the incidence
$\hat{S}_1$); otherwise the refraction is negative (indicated by
$\hat{S}_3'$ with respect to the incidence $\hat{S}_3$). When the
incidence is towards left, only positive refraction is possible,
referring to $\hat{S}_4'$ and $\hat{S}_4$.

\end{description}

\end{document}